\documentclass[twocolumn,aps,prd,epsf]{revtex4}
\pdfoutput=1
\usepackage[pdftex]{graphicx}
\usepackage{color}
\usepackage{amsmath}
\usepackage{braket}
\usepackage[T1]{fontenc}
\usepackage{url}
\usepackage[english]{babel}
\usepackage{graphicx}
\usepackage{dcolumn}
\usepackage{bm}
\usepackage{subfig}
\usepackage{float}
\usepackage{url}

\newcommand{\Tr}{\mbox{\rm Tr}}
\newcommand{\ReC}{\mbox{\rm Re}}

\newcommand\T{\rule{0pt}{2.6ex}}
\newcommand\B{\rule[-1.2ex]{0pt}{0pt}}

\usepackage{verbatim}

\begin{document}

\title{SU(3) gauge invariant  lattice QCD exploration of the dual superconductor picture in flux tube fusion,  in the dual gluon mass, and in
the dual Ginzburg-Landau parameters}
\author{
N. Cardoso}
\author{
M. Cardoso}
\author{
P. Bicudo}
\affiliation{CFTP, Departamento de F\'{\i}sica, Instituto Superior T\'{e}cnico,
Av. Rovisco Pais, 1049-001 Lisboa, Portugal}

\begin{abstract}
The colour fields, created by a static gluon-quark-antiquark system, are computed in quenched SU(3) lattice QCD, in a $24^3\times 48$ lattice at $\beta=6.2$ and $a=0.07261(85)\,fm$. We compute the hybrid Wilson Loop including the cases when the gluon and the antiquark are superposed, i. e., the quark-antiquark case and when the quark and antiquark are superposed, i. e., the gluon-gluon case. The Casimir scaling is investigated, in the two gluon glueball case the Casimir scaling is consistent with the formation of an adjoint string. Measuring the decay of the tail in the mid section of the flux tube for the two gluon glueball and for the quark-antiquark meson, we determine the penetration length and present a gauge invariant effective dual gluon mass of $0.905\pm0.163\,\text{GeV}$.
We also try to determine the coherence length comparing our results with the dual Ginzburg-Landau approach. With the penetration length and the possible coherence length we determine a putative Ginzburg-Landau dimensionless parameter, which is possibly consistent with a type II superconductor picture. These results are obtained at fixed quark-antiquark distance of 0.58 fm.
\end{abstract}
\maketitle

\section{Introduction}

\subsection{Dual superconductor picture}

Understanding how the confinement arises from QCD is a central problem of strong interaction physics.
In 1970's,  Nambu \cite{Nambu:1974zg}, 't Hooft \cite{'tHooft:1979uj} and Mandelstam \cite{Mandelstam:1974pi} proposed that quark confinement would be physically interpreted using the dual version of the superconductivity, the QCD vacuum state to behave like a magnetic superconductor. In the ordinary superconductor, Cooper-pair condensation leads to the Meissner effect, and
the magnetic flux is excluded or squeezed in a quasi-one-dimensional tube, the Abrikosov vortex, where the magnetic flux is quantized topologically.
This confinement mechanism has indeed been established in compact QED \cite{Polyakov:1975rs,Banks:1977cc,Smit:1989vg}.

\subsection{Flux tubes}

In the dual-superconductor picture for the QCD vacuum, the squeezing of the colour-electric flux between quarks is realized by the dual Meissner effect, as the result of condensation of colour-magnetic monopoles, which is the dual version of the electrical charged Cooper pair.  The QCD vacuum can be regarded as a dual version of a superconductor, based on the low-dimensionalization of the quantized flux between charges.
If magnetic monopoles are condensed in the vacuum, then the electric sources are confined by electric flux tubes, as magnetic charges would be confined by Abrikosov-Nielsen-Olesen (ANO) vortices \cite{Abrikosov:1956sx,Nielsen:1973cs,Cardoso:2006mf} in an ordinary superconductor (Meissner effect).
This is supported by experimental observations like Regge trajectories, \cite{Kaidalov:2001db}, and lattice simulations showing a linear quark-antiquark potential.
There is also evidence for the dual superconductor picture from numerical lattice QCD simulations \cite{Bali:1996dm,Gubarev:1999yp,Koma:2003gq,Chernodub:2005gz,Bornyakov:2003vx}.
Therefore, the flux tubes correspond to vortices and the flux between charges must be confined.

\begin{figure}[t!]
\begin{center}
    \includegraphics[width=9cm]{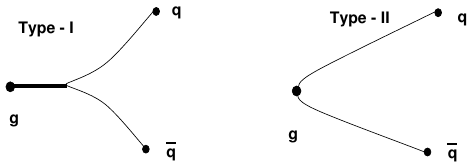}
    \caption{String attraction and fusion, and string repulsion, respectively in type I and II superconductors.}
    \label{superconductor}
\end{center}
\end{figure}

\subsection{Flux tube repulsion/attraction}

And analogy between established an analogy between the static potential and a type-II superconductor for the confinement in QCD  has been shown in Ref.  \cite{Cardoso:2007dc,Bicudo:2007xp} , and is illustrated in Fig. \ref{superconductor}, with repulsion of the fundamental strings and with the string tension of the first topological excitation of the string (the adjoint string) larger than the double of the fundamental string tension. In type-I superconductor the fundamental strings would be attracted and would fuse into an adjoint string. Ref. \cite{Cardoso:2009qt,Cardoso:2009kz} presented the colour fields for this system and the results are compatible with the formation of an adjoint string between two gluon glueball.

\subsection{ dual gluon mass}

The confinement of magnetic field lines in superconductors occurs through an Anderson-Higgs mechanism such that the photons of the electromagnetic field acquire a mass and are therefore exponentially damped in the superconductor.
In superconductors, the penetration length, $\lambda$, of the field in the London equation has a direct relation with an effective mass of the interaction particle fields, i. e., the photon.
Thus, by measuring the penetration length of the fields in the QCD vacuum, we may possibly estimate the mass of the dual gluon, if we further explore the analogy between QCD and superconductors. The dual gluon has been studied, \cite{Burdanov:1998tf,Jia:2005sp,Suzuki:2004uz,Suganuma:2004gq,Suganuma:2004ij,Suganuma:2003ds,Kumar:2004fj,Burdanov:2002ne}, however none of this studies have presented a value for the gauge invariant dual gluon mass in SU(3) lattice QCD.
In Table \ref{dual_gluon_mass_literature} and \ref{gluon_mass_literature}  we present some results found in the literature for the effective dual gluon mass and for the effective gluon mass, respectively.

\begin{table}
\begin{centering}
\begin{tabular}{|c|c|c|}
\hline
\T \B Mass, GeV  & Reference  & Estimation method\tabularnewline
\hline
\hline
\T \B $0.604$  & \cite{Baker:1991zh} & Dual QCD Langrangian\tabularnewline
\hline
\T \B $0.900$  & \cite{Suganuma:1997xk} & Lattice QCD, MA gauge\tabularnewline
\hline
\T \B $0.500$  & \cite{Suganuma:1997xk} & Lattice QCD, MA gauge\tabularnewline
\hline
\T \B $0.500$  & \cite{Tanaka:1999rj} & Lattice QCD, MA gauge\tabularnewline
\hline
\T \B $1.200$  & \cite{Suganuma:2000jh} & Lattice QCD, MA gauge\tabularnewline
\hline
\T \B $1.100$  & \cite{Suganuma:2002pm} & Lattice QCD, MA gauge\tabularnewline
\hline
\T \B $1.000$  & \cite{Suganuma:2003ds} & Lattice QCD, MA gauge\tabularnewline
\hline
\T \B $0.828$  & \cite{Kumar:2004fj} & QCD, Schwinger-Dyson equation\tabularnewline
\hline
\T \B $1.200$  & \cite{Suganuma:2004gq} & Lattice QCD, MA gauge\tabularnewline
\hline
\T \B $1.200$  & \cite{Suganuma:2004ij} & Lattice QCD, MA gauge\tabularnewline
\hline
\hline
\end{tabular}
\par\end{centering}
\caption{Estimates of the value of the effective dual gluon mass in the literature.}
\label{dual_gluon_mass_literature}
\end{table}

\begin{table}
\begin{centering}
\begin{tabular}{|c|c|c|}
\hline
\T \B Mass, GeV  & Reference  & Estimation method\tabularnewline
\hline
\hline
\T \B $0.800$  & \cite{Parisi:1980jy}  & $J/\psi\rightarrow\gamma X$ \tabularnewline
\hline
\T \B $0.500\pm0.200$  & \cite{Cornwall:1981zr} & Various \tabularnewline
\hline
\T \B $0.750$  & \cite{Spiridonov:1988md} & $\Pi_{\mu\nu}^{\text{e.m.}}$, $\Braket{\Tr G_{\mu\nu}^{2}}$\tabularnewline
\hline
\T \B $0.687-0.985$  & \cite{Donnachie:1988nj} & Pomeron parameters\tabularnewline
\hline
\T \B $0.800$  & \cite{Hancock:1992nj} & Pomeron slope\tabularnewline
\hline
\T \B $0.750$  & \cite{Nikolaev:1994vf} & Pomeron parameters\tabularnewline
\hline
\T \B $1.500_{-0.6}^{+1.2}$  & \cite{Field:1993fb} & PQCD at low scales (various)\tabularnewline
\hline
\T \B $1.460$  & \cite{Kogan:1994wf} & QCD vacuum energy, $\Braket{\Tr G_{\mu\nu}^{2}}$\tabularnewline
\hline
\T \B $10^{-10}-20\text{ MeV}$  & \cite{Yndurain:1995uq} & QCD potential\tabularnewline
\hline
\T \B $0.800$ & \cite{Szczepaniak:1995cw} & Coulomb gauge QCD Hamiltonian \tabularnewline
\hline
\T \B $0.570$  & \cite{Liu:1996gv} & $\Pi_{\mu\nu}^{\text{e.m.}}$, $\Braket{\Tr G_{\mu\nu}^{2}}$\tabularnewline
\hline
\T \B $0.470$  & \cite{Liu:1996gv} & Glueball current, $\Braket{\Tr G_{\mu\nu}^{2}}$\tabularnewline
\hline
\T \B $1.02\pm0.10$  & \cite{Leinweber:1998uu} & Lattice QCD, Landau gauge\tabularnewline
\hline
\T \B $0.800$ & \cite{LlanesEstrada:2000hj} & Coulomb gauge QCD Hamiltonian \tabularnewline
\hline
\T \B $0.800$ & \cite{LlanesEstrada:2000jw} & Coulomb gauge QCD Hamiltonian \tabularnewline
\hline
\T \B $0.721_{-0.009}^{+0.010}\,_{-0.068}^{+0.013}$  & \cite{Field:2001iu} & $J/\psi\rightarrow\gamma X$\tabularnewline
\hline
\T \B $1.180_{-0.06}^{+0.06}\,_{-0.28}^{+0.07}$  & \cite{Field:2001iu} & $\Upsilon\rightarrow\gamma X$\tabularnewline
\hline
\T \B $1.100$ & \cite{Silva:2004bv} & Lattice QCD, Landau gauge \tabularnewline
\hline
\T \B $0.800$ & \cite{LlanesEstrada:2005jf} & Coulomb gauge QCD model \tabularnewline
\hline
\T \B $0.650$ & \cite{Bicudo:2006sd} & BES $f_0(1810)$ \tabularnewline
\hline
\T \B $0.4\sim0.6$ & \cite{Iritani:2009mp} & Lattice QCD, Landau gauge \tabularnewline
\hline
\T \B $0.651(12)$ & \cite{Oliveira:2010xc} & Lattice QCD, Landau gauge \tabularnewline
\hline
\hline
\end{tabular}
\par\end{centering}
\caption{Estimates of the value of the effective gluon mass in the literature.}
\label{gluon_mass_literature}
\end{table}

\subsection{The Ginzburg-Landau model for superconductors}

Idealy, to establish the dual superconductor picture of confinement, one should not only determine the penetration length $\lambda$ but also the coherence length $\xi$ of the QCD vacuum, defined in the Ginzburg Landau equation. The Ginzburg Landau equation and the Ampere equation are given by
\begin{equation}
	\frac{\hbar^2}{2 m} ( \nabla - \frac{i q}{ \hbar c } \mathbf{A} )^2 \Psi + a \Psi - b |\Psi|^2 \Psi = 0\, ,
\end{equation}
\begin{equation}
	\nabla \times \nabla \times \mathbf{A} =
	\frac{q \hbar}{ 2 i m c } ( \Psi^* \nabla \Psi - \Psi \nabla \Psi^* - \frac{2 i q}{\hbar c} |\Psi|^2  \mathbf{A} )\, .
\end{equation}

Defining $a = \frac{\hbar^2}{2 m \xi^2}$, $\lambda = \sqrt{\frac{m c^2 b}{ q^2 a }}$, $\psi = \sqrt{\frac{b}{a}} \Psi$ and
$ \mathbf{a} = \frac{- q}{\hbar c} \mathbf{A}$, we arrive at the simplified equations
\begin{equation}
	\xi^2 ( \nabla + i \mathbf{a} )^2 \psi + \psi - b |\psi|^2 \psi = 0\,
\end{equation}

\begin{equation}
	\lambda^2 \nabla \times \nabla \times \mathbf{a} =
	\frac{\psi^*\nabla\psi-\psi\nabla\psi^*}{2 i} - |\psi|^2 \mathbf{a}\ .
\end{equation}

In the case of a vortex, we have, in cylinder coordinates $\mathbf{a}(\mathbf{r}) = a_{\varphi}(\rho) \hat{e}_{\varphi}$ and $\psi(\mathbf{r}) = u(\rho) e^{- i \varphi}$, so the equations are given by
\begin{equation}
	\xi^2 \big( \frac{1}{\rho} \frac{d}{d\rho} ( \rho \frac{d u}{d\rho} ) - \frac{u}{\rho^2}
	+ 2 \frac{a_\varphi}{\rho} u - a_\varphi^2 u \big) = u^3 - u\, ,
\end{equation}

\begin{equation}
	\frac{1}{\rho} \frac{d}{d\rho} ( \rho \frac{d u}{d\rho} ) - \frac{a_\varphi}{\rho^2} =
	\frac{1}{\lambda^2} u^2 ( \frac{1}{\rho} - a_\varphi )\, .
\end{equation}

\subsection{Summary}

In this paper, we further test the analogy between a dual superconductor and the pure gauge QCD flux tubes, including the first  exploration of the dual gluon mass in a SU(3) gauge independent lattice QCD.
In section II, we introduce the lattice QCD formulation.
We briefly review the Wilson loop for this system, which was used in Bicudo et al. \cite{Bicudo:2007xp},  Cardoso et al. \cite{Cardoso:2007dc} and Cardoso et al. \cite{Cardoso:2009kz}, and show how we compute the colour fields and as well as the lagrangian and energy densities distributions.
In section III, the numerical results are shown.
We present the results for the color field profiles in the mid flux tube section for the static hybrid gluon-quark-antiquark system, in a U shape geometry.
A detailed study of the Casimir scaling is done.
Measuring the penetration length of the tail in the mid flux tube, we also present a value for the effective dual gluon mass. We finally try to measure the coherence length and the Ginzburg-Landau dimensionless parameter.
Finally, we present our conclusion in section IV.

\section{The Wilson Loops and Colour Fields}

\begin{figure}[h]
\begin{centering}
    \subfloat[\label{loop0}]{
\begin{centering}
    \includegraphics[width=6cm]{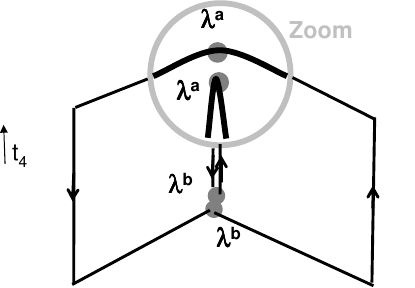}
\par\end{centering}}
\\
    \subfloat[\label{loop1}]{
\begin{centering}
    \includegraphics[width=7cm]{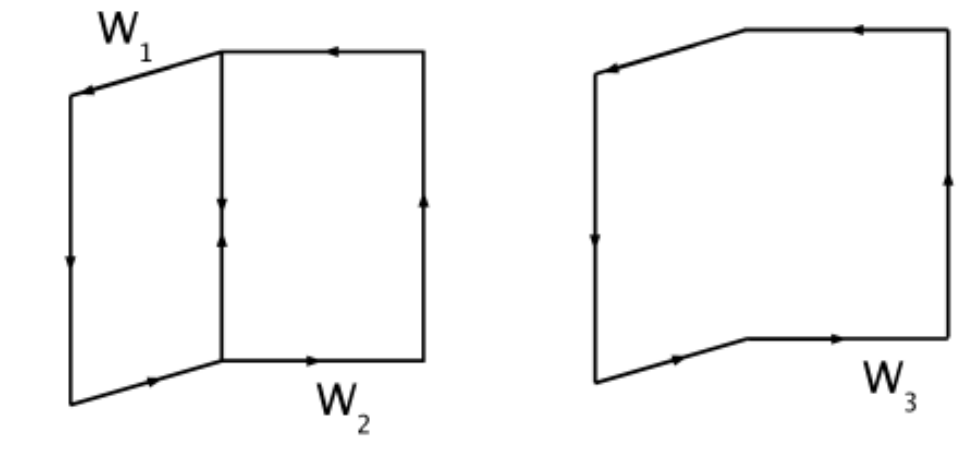}
\par\end{centering}}
\par\end{centering}
    \caption{\protect\subref{loop0} Wilson loop for the $gq\overline{q}$ and equivalent position of the static antiquark, gluon, and quark. \protect\subref{loop1} Simple Wilson loops that make the $gq\overline{q}$ Wilson loop.}
    \label{loop}
\end{figure}

The Wilson loop for the static hybrid gluon-quark-antiquark was deducted in \cite{Bicudo:2007xp,Cardoso:2007dc,Cardoso:2009qt,Cardoso:2009kz}, therefore we only present the fundamental expressions. The Wilson loop for this system is given by
\begin{equation}
    W_{gq\overline{q}}=W_1 W_2 - \frac{1}{3}W_3
\label{wloopqqg}
\end{equation}
where $W_1$, $W_2$ and $W_3$ are the simple Wilson loops shown in Fig. \ref{loop}.
Importantly, for the study of the Casimir Scaling,
when $r_1=0$, $W_1=3$ and $W_2=W_3$, the operator reduces to the mesonic Wilson loop
and when $\mu=\nu$ and $r_1=r_2=r$, $W_2=W_1^{\dagger}$ and $W_3=3$, $W_{gq\overline{q}}$ reduces
to $W_{gq\overline{q}}(r,r,t)=|W(r,t)|^2-1$, that is the Wilson loop in the adjoint representation
used to compute the potential between two static gluons.

In order to improve the signal to noise ratio of the Wilson loop, we use the APE smearing defined by
\begin{eqnarray}
    U_{\mu}\left(s\right) & \rightarrow & P_{SU(3)}\frac{1}{1+6w}\Big(U_{\mu}\left(s\right)\nonumber \\
    & & + w \sum_{\mu\neq\nu}U_{\nu}\left(s\right) U_{\mu}\left(s+\nu\right)U_{\nu}^{\dagger}\left(s+\mu\right)\Big)\ ,
\end{eqnarray}
with $w = 0.2$ and iterate this procedure 25 times in the spatial direction.

To achieve better accuracy in the flux tube, we apply the  hypercubic blocking (HYP) in the time direction, \cite{Hasenfratz:2001hp}, with
\begin{equation}
\alpha_1=0.75, \quad \alpha_2=0.6, \quad \alpha_3=0.3\ .
\end{equation}

We obtain the chromoelectric and chromomagnetic fields on the lattice, by using,
\begin{equation}
    \Braket{E^2_i}= \Braket{P_{0i}}-\frac{\Braket{W\,P_{0i}}}{\Braket{W}}
\end{equation}
and,
\begin{equation}
    \Braket{B^2_i}= \frac{\Braket{W\,P_{jk}}}{\Braket{W}}-\Braket{P_{jk}}
\end{equation}
where the $jk$ indices of the plaquette complement the index $i$ of the magnetic field,
and where the plaquette is given by
\begin{equation}
P_{\mu\nu}\left(s\right)=1 - \frac{1}{3} \ReC\,\Tr\left[ U_{\mu}(s) U_{\nu}(s+\mu) U_{\mu}^\dagger(s+\nu) U_{\nu}^\dagger(s) \right]\ .
\end{equation}
The energy ($\mathcal{H}$) and lagrangian ($\mathcal{L}$) densities are given by
\begin{equation}
    \mathcal{H} = \frac{1}{2}\left( \Braket{E^2} + \Braket{B^2}\right)\ ,
    \label{energy_density}
\end{equation}
\begin{equation}
    \mathcal{L} = \frac{1}{2}\left( \Braket{E^2} - \Braket{B^2}\right)\ .
    \label{lagrangian_density}
\end{equation}
Notice that we only apply the smearing technique to the Wilson loop.

\section{Results}

Here we present the results of our simulations with 286 $24^3 \times 48$, $\beta = 6.2$ quenched configurations generated with the version 6 of the MILC code \cite{MILC}, via a combination of Cabbibo-Mariani and overrelaxed updates.
The results are presented in lattice spacing units of $a$, with $a=0.07261(85)\,fm$ or $a^{-1}=2718\,\pm\, 32\,MeV$.

\begin{figure}[h]
\begin{centering}
    \subfloat[U shape geometry.\label{fig:shapeU}]{
\begin{centering}
    \includegraphics[height=3.5cm]{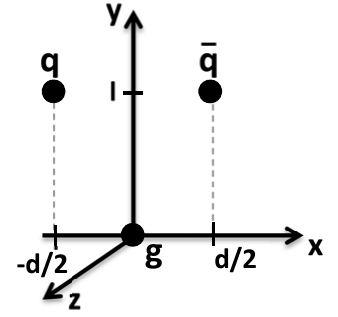}
\par\end{centering}}
    \subfloat[L shape geometry.\label{fig:shapeL}]{
\begin{centering}
    \includegraphics[height=3.5cm]{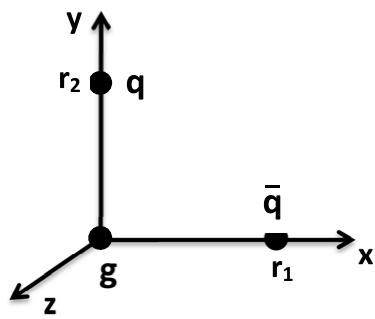}
\par\end{centering}}
\par\end{centering}
    \caption{gluon-quark-antiquark geometries, U and L shapes.}
    \label{shape}
\end{figure}

In this work two geometries for the hybrid system, gluon-quark-antiquark, are investigated: a U shape and a L shape geometry, both defined in Fig. \ref{shape}. Note that, in the L shape geometry only one case is studied here, the case when the gluon and the antiquark are superposed, $r_1=0$, this case corresponds to the quark-antiquark case. This particular case is relevant to study the Casimir scaling.

\begin{figure}
\begin{centering}
    \subfloat[$\mathcal{L}$ at $y=4$ and $z=0$\label{qqg_U_Act_ape_hyp}]{
\begin{centering}
    \includegraphics[width=8cm]{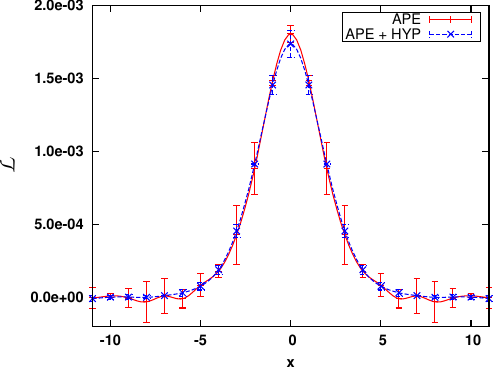}
\par\end{centering}}

    \subfloat[$\mathcal{L}$ with APE at $z=0$.\label{cfield_QQG_APE_U_Sim_d_0_l_8_Act}]{
\begin{centering}
    \includegraphics[height=3.4cm]{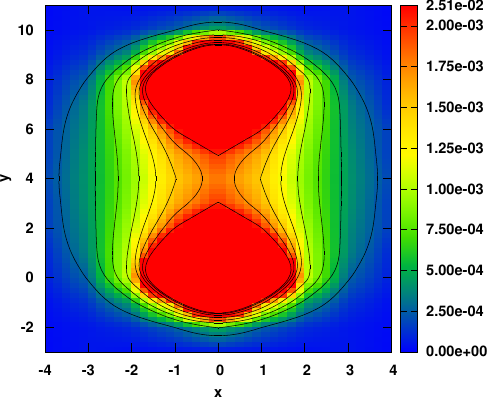}
\par\end{centering}}
    \subfloat[$\mathcal{L}$ with APE and HYP at $z=0$.\label{cfield_QQG_HT_U_Sim_d_0_l_8_Act}]{
\begin{centering}
    \includegraphics[height=3.4cm]{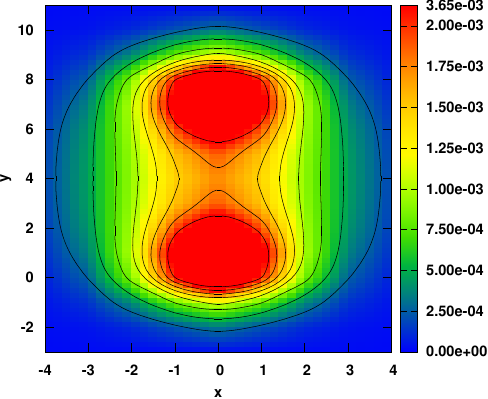}
\par\end{centering}}
\par\end{centering}
\caption{Comparison between results for lagrangian density with APE smearing only in space and for lagrangian density with APE smearing in space and Hypercubic Blocking (HYP) in Time for the U geometry with $d=0$ and $l=8$. In \protect\subref{qqg_U_Act_ape_hyp}, the lines were drawn for convenience and therefore do not represent results from any kind of interpolation.}
\label{APE_HYP}
\end{figure}

The use of the APE (in space) and HYP (in time) smearing allows us to have better results for the flux tube and reduce large plaquette fluctuations. Notice in Fig. \ref{APE_HYP} the improvement of the signal in the tail of the flux tube profile.

\begin{figure*}
\begin{centering}
    \subfloat[$\Braket{E^2}$\label{qqg_U_ape_hyp_E_y=4}]{
\begin{centering}
    \includegraphics[height=6cm]{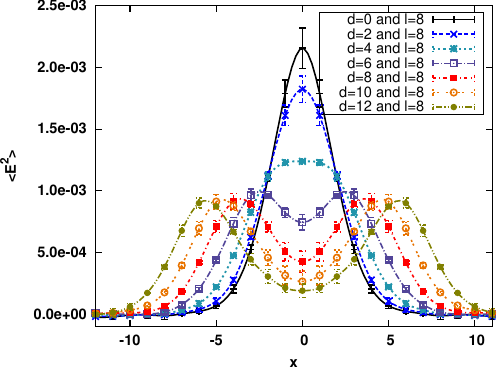}
\par\end{centering}}
    \subfloat[$-\Braket{B^2}$\label{qqg_U_ape_B_Act_y=4}]{
\begin{centering}
    \includegraphics[height=6cm]{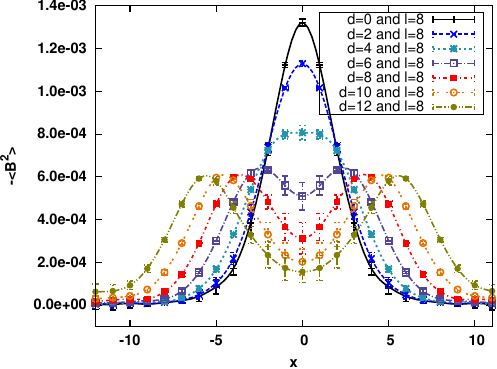}
\par\end{centering}}

    \subfloat[$\mathcal{L}$\label{qqg_U_ape_hyp_Act_y=4}]{
\begin{centering}
    \includegraphics[height=6cm]{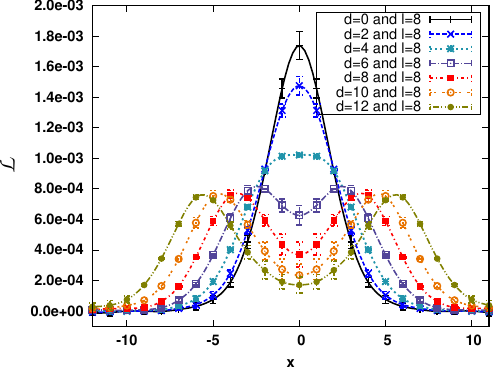}
\par\end{centering}}
    \subfloat[$\mathcal{H}$\label{qqg_U_ape_hyp_Energ_y=4}]{
\begin{centering}
    \includegraphics[height=6cm]{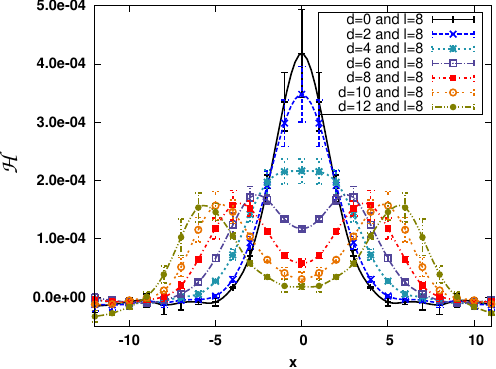}
\par\end{centering}}
\par\end{centering}
    \caption{Results for the U geometry at $y=4$ and $z=0$.}
    \label{qqg_U_Sim_profile}
\end{figure*}

\begin{figure*}
\begin{centering}
    \subfloat[$r=y$ at $x=0$ and $z=0$\label{casimir_xy_Energ_x=0}]{
\begin{centering}
    \includegraphics[height=6cm]{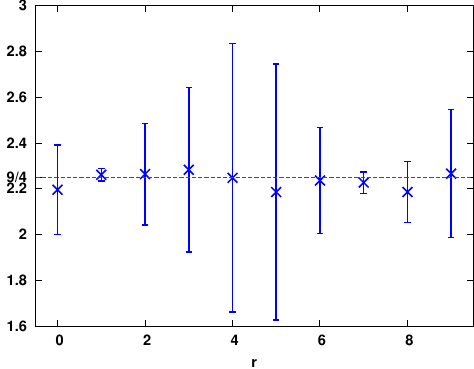}
\par\end{centering}}
    \subfloat[$r=(x,z)$ at $y=4$\label{casimir_xz_Energ}]{
\begin{centering}
    \includegraphics[height=6cm]{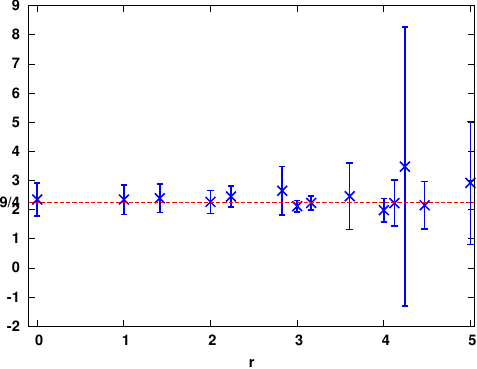}
\par\end{centering}}
\par\end{centering}
    \caption{Results for the two gluon glueball ($d = 0$ and $l = 8$, U
geometry) energy density over the meson ($r_1 = 0$ and
$r_2 = 8$, L geometry) energy density at $x=0$ and Casimir
scaling factor of $9/4$ (broken line).}
    \label{casimir}
\end{figure*}

\subsection{U profiles and Casimir scaling}

In Fig. \ref{qqg_U_Sim_profile} we present the profiles for the U geometry with $l=8$ and $d$ between 0 and 12 at $y=4$.
We can see the stretching and partial splitting of the flux tube in the equatorial plane ($y = 4$) between the quark and antiquark and $z=0$. When $d=0$, the quark and the antiquark are superposed and this corresponds to the two gluon glueball case.

For $d=2$ and 4 at $y=4$, the separation between the two flux tube is not visible, this is due to the overlap in the tails of the flux tube which contributes for the total field and, for large separations, the tails of the flux tubes contribute to a non zero field at $x=0$.

We measure the quotient between the energy density of the two gluon glueball system and of the meson system, in the
mediator plane between the two particles ($x = 0$).
The results are shown in Fig. \ref{casimir}.
In Fig. \ref{casimir_xy_Energ_x=0} we present the results for $r=y$ at $x=z=0$ and in Fig. \ref{casimir_xz_Energ} we present the results for $r=(x,z)$ at $y=4$.
We make a constant fit to the data in Fig. \ref{casimir}, the result for Fig. \ref{casimir_xy_Energ_x=0} is $2.25096 \, \pm \, 0.0244972$ and for Fig. \ref{casimir_xz_Energ} is $ 2.23591 \, \pm \,0.0598732$.
As it can be seen, these results are consistent with the Casimir scaling, with a factor of $9/4$ between the energy density in the glueball and
in the meson.
This corresponds to the formation of an adjoint string between the two gluons, and in the quark-antiquark case we have a fundamental string. The points outside the field, i.e., as the field approaches to zero, were discarded.

\subsection{Dual Gluon Mass}

In common superconductivity the magnetic field decays with $B \sim e^{-r/\lambda_L}$ and this could be interpreted in terms of an effective mass for the photon $m_\gamma = 1 / \lambda_L$. Some studies have pointed a similar behaviour in QCD, \cite{DiGiacomo:1992df,Baker:1984qh,Bali:1998de,Jia:2005sp}, as in a superconductor, the chromoelectric field decays as $\sim e^{-r/\lambda_L}$ as they depart from the singular vortex flux tube, where $r$ stands for the distance away from the centre of the flux tube. This is consistent with the dual superconductor picture.

We measure the decay of the mid flux tube section in the quark-antiquark and the two gluon glueball.
We tried to fit our data by a gaussian and by a exponential, but the $\chi^2/dof$ clearly indicated that the decay is exponential rather than the gaussian.
This result reinforces the correspondence between the QCD vacuum and the Dual Superconductor.
Baker et al. \cite{Baker:1984qh} presented as solution to the electric and magnetic colour flux tube the modified Bessel functions. We also fitted our data with the modified Bessel function of zero order, $K_0\left( \mu\, r\right)$.

So in our case, the mean value of the squared fields should decay as $e^{-2\, \mu\, r}$ or like $K_{0}^{2}\left(\mu\, r\right)$, where $\mu$ should be the effective dual gluon mass.
However, since we have a flux tube and not a wall, i. e., the case study is not planar, the correct fit is the modified Bessel function of zero order, suitable for cylindrical flux tubes.

Since $\Braket{E^2}$, $\Braket{B^2}$, $\mathcal{H}$ and $\mathcal{L}$ are all gauge invariant, we can compute a value for
the effective dual gluon mass that is gauge independent by studying the decay of one of these quantities in the middle of the flux tube, between the sources. Fitting the chromoelectric field and the lagrangian density section in the mid distance of the flux tube of the meson and the two gluon glueball, we obtain the results presented in Table \ref{tab_dual_gluon_mass} for the effective dual gluon mass. The results are compatible with an identical decay of the Lagrangian density in both cases and are of the order of $ \sim 1\, \text{GeV} $.

\begin{table}
\begin{centering}
\begin{tabular}{|c|c|c|c|c|}
\cline{2-5}
\multicolumn{1}{c|}{} & \multicolumn{2}{c|}{\T \B $a\, e^{-2\, \mu\, r}$} & \multicolumn{2}{c|}{$a\, K_{0}^{2}\left(\mu\, r\right)$}\tabularnewline
\cline{2-5}
\multicolumn{1}{c|}{} & \T \B $\mu\ \left(\text{GeV}\right)$ & $\chi^{2}/dof$ & $\mu\ \left(\text{GeV}\right)$ & $\chi^{2}/dof$\tabularnewline
\hline
\hline

\T \B $E_{(1a)}^{2}\left(r\right)$ & $1.170\pm0.228$ & $1.069$ & $0.805\pm0.287$ & $1.827$\tabularnewline
\hline
\T \B $\mathcal{L}_{(1a)}\left(r\right)$ & $1.170\pm0.119$ & $0.512$ & $0.865\pm0.188$ & $1.203$\tabularnewline
\hline
\T \B $E_{(2a)}^{2}\left(r\right)$ & $1.231\pm0.286$ & $1.547$ & $0.881\pm0.334$ & $2.084$\tabularnewline
\hline
\hline

\T \B $E_{(1b)}^{2}\left(r\right)$ & $1.210\pm0.056$ & $0.887$ & $0.897\pm0.085$ & $1.185$\tabularnewline
\hline
\T \B $\mathcal{L}_{(1b)}\left(r\right)$ & $1.208\pm0.068$ & $0.560$ & $0.909\pm0.099$ & $0.909$\tabularnewline
\hline
\T \B $E_{(2b)}^{2}\left(r\right)$ & $1.298\pm0.098$ & $1.272$ & $1.027\pm0.088$ & $1.331$\tabularnewline
\hline
\T \B $\mathcal{L}_{(2b)}\left(r\right)$ & $1.216\pm0.047$ & $1.183$ & $0.949\pm0.062$ & $1.215$\tabularnewline
\hline
\hline
\end{tabular}
\par\end{centering}
\caption{Results for the dual gluon mass, where (1a) is for the two gluon glueball and (2a) for the quark-antiquark cases at $y=4$ and $z=0$ with $r=x>2$ and $z=0$, and (1b) is for the two gluon glueball and (2b) for the quark-antiquark cases at $y=4$ with $r=(x,z)>2$.}
\label{tab_dual_gluon_mass}
\end{table}

\subsection{Comparing with the solution of the Ginzburg Landau Equations}

\begin{table*}
\begin{centering}
\begin{tabular}{|c|c|c|c|c|c|}
\hline
\T \B & $\lambda$ (a) & $\xi$ (a) & $\chi^{2}/dof$ & $\kappa$\tabularnewline
\hline
\hline
\T \B $\Braket{E^2_{gg}}$ & $3.031\pm0.261$ & $2.294\pm0.011$ & $0.425$ & $1.321\pm0.114$\tabularnewline
\hline
\T \B $\mathcal{L}_{gg}$ & $2.991\pm0.295$ & $2.356\pm0.010$ & $0.403$ & $1.270\pm0.125$\tabularnewline
\hline
\T \B $\Braket{E^2_{q\overline{q}}}$ & $2.773\pm0.240$& $2.277\pm0.053$ & $1.110$ & $1.218\pm0.109$\tabularnewline
\hline
\T \B $\mathcal{L}_{q\overline{q}}$ & $2.865\pm0.193$ & $2.371\pm0.014$ & $0.268$ & $1.208\pm0.082$\tabularnewline
\hline
\hline
\end{tabular}
\par\end{centering}
\caption{Results obtained by fitting the numeric solution of the Ginzburg-Landau and Ampere equations to the colour fields and lagrangian density data with $\lambda$ fixed. $gg$ - two gluon glueball results, $q\overline{q}$ - quark-antiquark results in XZ plane at $y=4$.}
\label{xi_and_lambda}
\end{table*}

The failure of fitting the entire set of the chromoelectric in the mid flux tube data, with the exponential or the modified Bessel function only, hints at the existence of another scale, in addition to the dual gluon mass, $\mu=\lambda^{-1}$. 

The extraction of this second scale, the coherence length, $\xi$, can be attempted using the effective dual Ginzburg-Landau approach. The ratio between the penetration length and the coherence length,
\begin{equation}
\kappa=\frac{\lambda}{\xi}
\end{equation}
is the famous Ginzburg-Landau dimensionless parameter used to determine whether a superconductor is type I or type II.

However we are not able to obtain a good fit of the two Ginzburd-Landau parameters due to redundancy. Perhaps this happens because in the Ginzburg-Landau equations the total magnetic flux is quantized by the scale of the electron charge and the plank constant $e \over h$. In QCD the quantization of the electric field flux has not yet been established and we miss a constraint. 

\begin{figure}[h]
\begin{centering}
    \subfloat[Two gluon glueball case.\label{fig:U_E_r}]{
\begin{centering}
    \includegraphics[height=5.5cm]{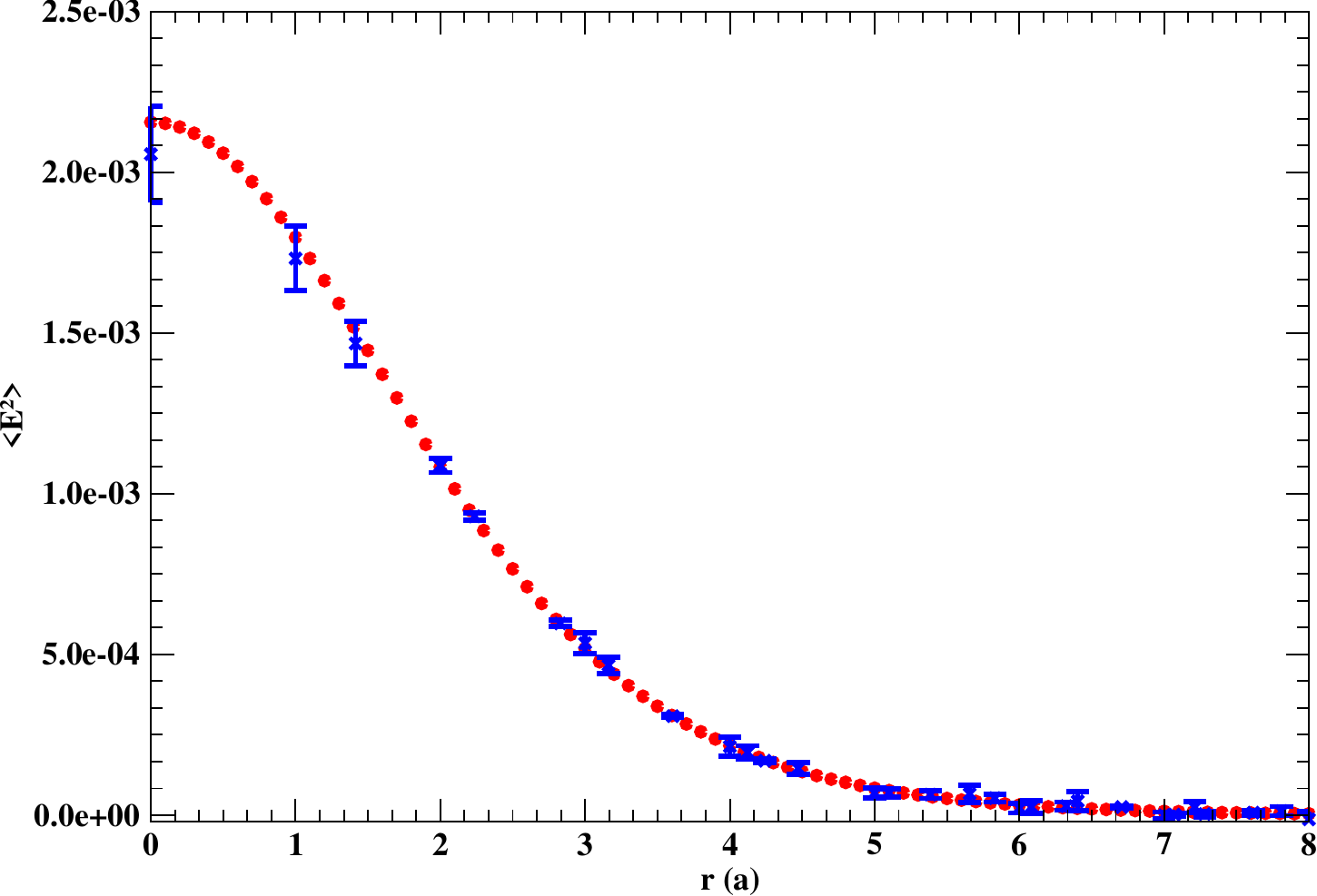}
\par\end{centering}}

    \subfloat[Quark-antiquark case.\label{fig:L_E_r}]{
\begin{centering}
    \includegraphics[height=5.5cm]{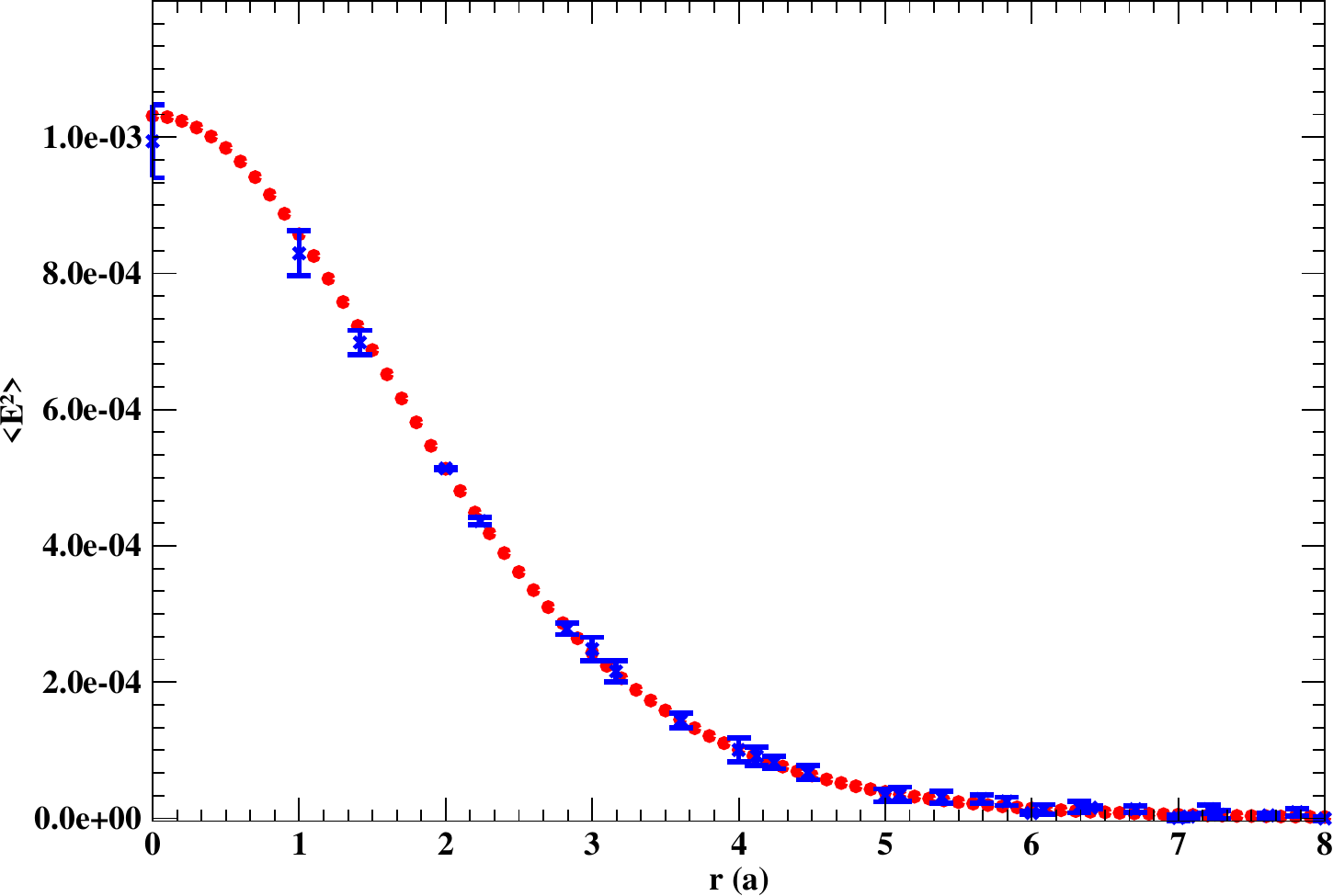}
\par\end{centering}}
\par\end{centering}
    \caption{Fit to the chromoelectric field data using the numeric solution of the Ginzburg-Landau and Ampere equations, red points. The blue points are the results for chromoelectric field.}
    \label{fig_GLE_FIT}
\end{figure}

Thus we opt for solving these equations with $\lambda$ fixed, obtained from the modified Bessel function of zero order fit in Table \ref{tab_dual_gluon_mass}. Fitting the numerical result to the the results of the colour fields and the lagrangian density in the middle of the flux tube, in XZ plane, we obtain the results in Table \ref{xi_and_lambda} and Fig. \ref{fig_GLE_FIT}. This leads to a Ginzburg-Landau parameter, $\kappa$, greater than $2^{-1/2}$, which corresponds to a type II superconductor.

\section{Conclusions}

We present the first gauge invariant result for the effective dual gluon mass in a pure gauge SU(3) QCD lattice. This an exploratory result only, since we considered only a charge-anticharge distance of 8 lattice spacings, and results are expected to depend on this distance. 

A detailed study of the Casimir scaling, comparing the triplet and the octet representations, is presented. The results are consistent with the formation of an adjoint string between the two gluon, in agreement with the Casimir Scaling measured by Bali \cite{Bali:2000un}, with a factor of $9/4$.

We also study the decay of the fields perpendicular to the flux tube, and obtain an exponential decay consistent with the dual superconductor model of confinement. By considering the decay factor of the mid distance tube flux, we obtain a gauge invariant result of $ \sim 1 \text{ GeV} $ for the effective dual gluon mass and this value is consistent with previous analytical theoretical calculations and independent phenomenological estimates for the effective dual gluon mass and gluon mass found in the literature.

Applying the Ginzburg Landau and the Ampere equations, we are able to perform a complete fit to the colour fields in the mid flux tube and determine the coherence length parameter. Notice we needed to maintain the penetration length fixed and determined from the fit of the tail of the profile with a modified Bessel function of zero order. With these two parameters, we determine the Ginzburg-Landau dimensionless parameter, $\kappa$, which indicates a type II superconductor region as detailed in Table \ref{xi_and_lambda}. Again this result is obtained for an inter-charge distance of 8 lattice spacings.

However, SU(3) QCD is not completely equivalent to a dual superconductor. In one hand the total flux is not quantized with a universal law as in a superconductor, and this prevents us from a completely independent fit of the Ginzburg-Landau parameters.  Moreover we find that all six components $E_i^2$ and $B_i^2$ have similar flux tube profiles, while in a superconductor only the component of the magnetic field $\mathbf{B}$ parallel to the flux tube has such a profile. The dominant component of the chromoelectric field, $E_y^2$, parallel to the flux tube, is about two times the $E_x^2$ and $E_z^2$ components. In the chromomagnetic field the $B_x^2$ and $B_z^2$ are dominant and are about is about 1.5 times the $B_y^2$ component in the flux tube. The dominant component of the chromoelectric field is about 3.3 times the $|B_y^2|$ and 2.2 the $|B_x^2|$ and $|B_z^2|$ in this region. This may suggest in QCD the existence of a gluon mass, identical to the dual gluon mass.

Combining all $E_i^2$ and $B_i^2$ components we find the result for the putative dual gluon and gluon masses of $0.905\pm0.163\,\text{GeV}$.

\acknowledgments
We acknowledge discussions on superconductors with Pedro Sacramento.
This work was financed by the FCT contracts POCI/FP/81933/2007 and CERN/FP/83582/2008.
We thank Orlando Oliveira for useful discussions and for sharing gauge field configurations.

\bibliographystyle{apsrev4-1}
\bibliography{paper3}

\end{document}